# Heterogeneous Wettability Alters Methane Migration and Leakage in Shallow Aquifers

Sabber Khandoozi[a,*], Siddharth Gautam[b], Craig Dietsch[a], Muhammad Sahimi[c], David Cole[b], Mohamad Reza Soltanian[a,d,**]

[a]Department of Geosciences, University of Cincinnati, Cincinnati, OH, USA

[b]School of Earth Sciences, The Ohio State University, Columbus, OH, 43210, USA

[c]Mork Family Department of Chemical Engineering and Materials Science, University of Southern California, USA

[d]Department of Environmental Engineering, University of Cincinnati, Cincinnati, OH, USA

**Corresponding authors:** [*]khandosr@uc.edu

[**]soltanma@uc.edu

**Abstract**

Capillary heterogeneity is increasingly recognized as a first-order control on gas plume migration and trapping in aquifers and storage formations. We show that spatial variability in the water–methane contact angle, set by mineralogy and salinity, reshapes capillary entry pressures and, in turn, migration pathways. Using molecular dynamics simulation, we estimate contact angles on quartz and kaolinite under fresh and saline conditions and embed these results in continuum-scale multiphase flow simulations via a contact-angle–informed Leverett *J*-function, mapping wettability directly onto the flow properties at continuum scale. Accounting for contact-angle heterogeneity changes methane behavior: mobile and residually trapped methane in aquifers decrease by up to 10%, while leakage to the atmosphere increases by as much as 20%. The magnitude of this effect is scaled with the permeability contrast, leakage rate, salinity, and facies proportions. By coupling molecular-scale wettability to continuum-scale flow and transport, the cross-scale framework provides a more physically grounded basis for groundwater protection and risk assessments, and yields more reliable emissions estimates. The approach can be generalized to other subsurface gas transport problems including hydrogen and carbon dioxide storage as well as to natural releases such as methane from permafrost thaw.

**Keywords:**

Fugitive methane leakage; Capillary heterogeneity; Heterogeneous wettability; Molecular dynamics; Atmospheric methane emissions

**Synopsis**:

Contact angles, computed by molecular dynamics simulations, link mineral- and salinity-dependent wettability to methane migration. Integrating them into continuum-scale models reduces methane trapping and increases atmospheric emissions, informing groundwater protection and emission risk assessments.

## Graphical Abstract

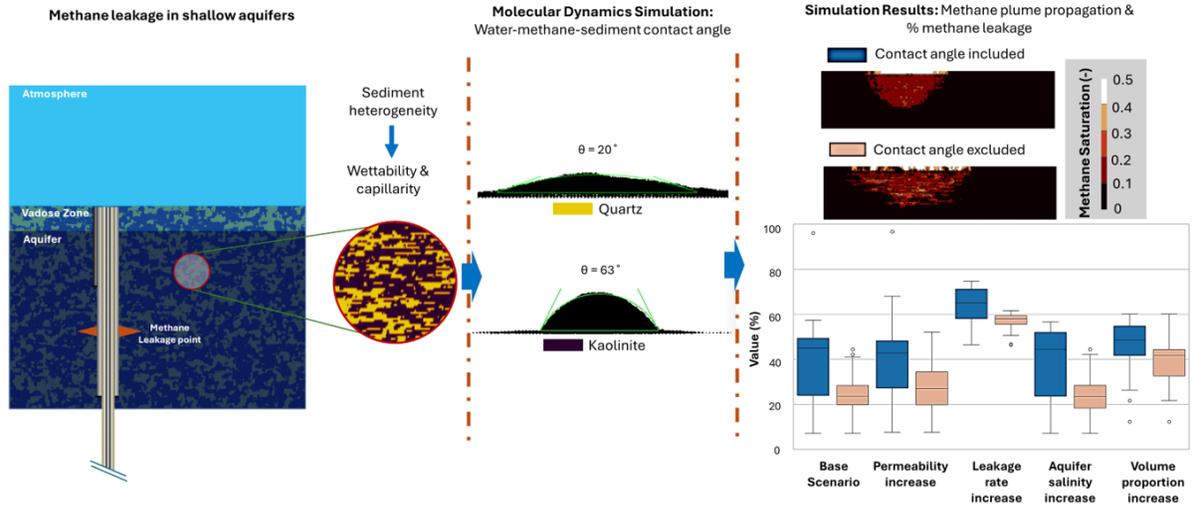

# 1 Introduction

Methane ($CH_4$) leakage from subsurface sources into shallow aquifers and the atmosphere is a significant environmental concern [1–4]. Fugitive $CH_4$ may originate from compromised wells, natural faults, fractures, or surface activities, and is further complicated by naturally occurring biogenic $CH_4$ [5–11]. As a potent greenhouse gas, $CH_4$ poses risks to groundwater quality, climate, and public safety, making it essential to predict its migration pathways [12–15]. Recent work highlights the strong influence of small-scale geologic features, with capillary heterogeneity identified as a primary controlling factor for plume migration and trapping [16,17]. Neglecting capillary pressure variability biases predictions, often overstating atmospheric emissions and understating aquifer retention [16]. Building on this foundation, we investigate how mineral- and salinity-dependent wettability, expressed as water–$CH_4$ contact angle heterogeneity, further reshapes leakage pathways when integrated across molecular to aquifer scales.

Field studies and numerical models have demonstrated that small-scale structural features and variations in capillary properties can redirect plume pathways, alter trapping efficiency, and change leakage outcomes [16–18]. Capillary heterogeneity has been recognized as a critical control on gas migration [16,19–25]. However, field studies have shown that neglecting this factor may introduce systematic bias into gas plume migration predictions [19–21,25,26]. Beyond porosity- and permeability-based scaling of capillary pressure, wettability represents an underexplored dimension of heterogeneity with equally important consequences for plume behavior.

Conventional capillary pressure scaling in $CH_4$ leakage studies and in related problems (e.g., $CO_2$ and $H_2$ storage) typically relies on spatial variation in porosity and permeability while holding interfacial properties fixed [16,27–32]. This approach overlooks the fact that wettability depends on mineralogy and water chemistry (e.g., salinity), which changes local capillary entry pressures and, in turn, migration pathways, but is rarely included in leakage modeling.

Direct measurements of contact angle (e.g., sessile/pendant drop) are valuable but face well-known challenges such as surface contamination, unwanted roughness, and damage due to smoothing during core preparation. As a result, the measured contact angle may not truly represent the natural surface conditions. These challenges are particularly acute in unconsolidated shallow aquifers [33,34]. Molecular dynamics (MD) simulation provides a complementary approach by resolving atomic-scale interactions at mineral–fluid–fluid interfaces. Prior studies show good agreement between MD-computed and laboratory-measured contact angles across a range of rock–fluid systems [35–39]. In this way, MD simulation can supply facies-specific wettability inputs that can be transferred to continuum-scale models through the Leverett *J*-function, thereby making an explicit link between nanoscale surface physics and continuum-scale multiphase flow.

Here, we explore the link between microscale wettability and macroscale leakage behavior by integrating contact angles, computed from MD simulations as functions of mineral type and salinity, into continuum-scale multiphase flow simulations of $CH_4$ leakage along a shallow aquifer wellbore. This integration is achieved using a contact-angle-informed Leverett *J*-function, and evaluate how explicitly representing heterogeneous wettability influences $CH_4$ migration, trapping, and leakage. We further identify the conditions under which this representation matters most by varying salinity, permeability contrast, leakage rate, and facies proportions within geologically plausible architectures.

## 2 Materials and methods

Integrating MD simulation-computed, mineral- and salinity-dependent contact angles with continuum-scale multiphase flow simulations were performed in two stages. First, MD simulations provided facies-specific contact angles under varying salinity conditions. Second, the computed contact angles were incorporated into continuum-scale models to refine capillary pressure scaling. The resulting simulations were used to quantify $CH_4$ leakage to the atmosphere and to identify the conditions where quantifying contact angle variability is most critical for accurate predictions. Computational costs for both MD simulations and continuum-scale simulations are detailed in Supporting Information, **Section S1**.

All MD simulations were performed with the LAMMPS package [40] with MPI parallelization and GPU acceleration [41–46]. Complete details of the force fields, equilibration protocols, and analysis procedures along with comparison with prior lab results are provided in Supporting Information,

**Section S2**. Although shallow aquifers contain multiple mineral phases, previous studies indicated that small-scale mineralogical heterogeneity exerts limited influence on solute transport, with plume migration and spreading governed primarily by larger-scale hydrogeologic structure [24]. Based on the evidence, sediment variability was simplified into two principal mono-mineralic facies: coarse grain (CG) and fine grain (FG) facies (i.e., bodies of sediment) [47]. Quartz was selected as the representative mineral for CG facies, while kaolinite-coated surfaces represented FG facies.

Quartz exists in various polymorphic forms, each characterized by a unique atomic arrangement, depending on the pressure and temperature conditions of the porous medium [1]. Among these, α-quartz with a (0001) surface is selected due to its stability under the target simulation conditions [1,2]. Since α-quartz crystallizes in a hexagonal structure, and the simulation box requires a rectangular geometry, modifications to the crystal lattice are necessary. To achieve this, the α-quartz unit cell [3] is duplicated along the x and y directions, then trimmed along specific crystallographic planes to ensure two-dimensional periodicity and neutral charge of substrate. An orthogonal unit cell is then constructed and replicated along the *x*, *y*, and *z* axes, maintaining atomic stoichiometry and bond connectivity throughout the structure [4]. A Q2-cut is extracted from the generated α-quartz surface and fully hydroxylated to represent realistic aqueous conditions. For Q2, the silicon atom is bonded to two bridging oxygen atoms and two hydroxyl (OH) groups. This is a less stable, highly reactive silanol group, found on the surface of silica, with the structure $(\equiv Si-O-)_2 Si(OH)_2$. In contrast, kaolinite ($Al_2Si_2O_5(OH)_4$) is a non-swelling, 1:1 type clay mineral composed of one tetrahedral silica (T) sheet and one octahedral alumina (O) sheet [5]. Unlike smectites, kaolinite has negligible layer charge, and its interlayers are stabilized by hydrogen bonding between hydroxyl groups on the alumina sheet and basal oxygen atoms on the silica sheet, forming a rigid TO–TO stacking with minimal interlayer expansion. The kaolinite unit cell is first extended in the *x* and *y* directions, then replicated three times along the *z*-axis to generate the simulation domain.

Water–$CH_4$ contact angles on these minerals were estimated using established MD simulation protocols that have been shown to reliably reproduce laboratory-scale measurements [38–40]. In this study, a semi-cylindrical water droplet was placed on the mineral substrate in the presence of $CH_4$ and equilibrated to steady geometry (Supporting Information, **Figure S1**). In the simulation box, periodic boundary conditions are applied in the x- and y-directions to simulate an infinite system, ensuring that when a particle exits one boundary, it seamlessly re-enters from the opposite side. In contrast, nonperiodic boundary conditions are enforced in the z-direction, where atoms may be reflected, absorbed, or constrained within the simulation box. To minimize boundary effects, a reflective wall is placed at the top of the simulation box, and the box height was selected to be sufficiently large to prevent any influence of the walls on the contact angle measurements. The size of the simulation box is about 120 nm in z-direction, including the substrate. The contact angle was measured at the three-phase boundary between the water droplet, mineral surface, and $CH_4$ phase to be used in continuum-scale simulations.

Continuum-scale simulations were conducted using the GEM module of Computer Modelling Group (CMG, Version 2024.30). The range of the properties used in continuum-scale simulations is based on prior field studies and numerical simulations in $CH_4$ leakage in unconfined shallow aquifers [17,23]. The model domain was divided into three sections representing an unconfined shallow aquifer with an overlying vadose zone and atmosphere (Supporting Information, **Section S3**). Initial saturations were specified as 100% water in the aquifer, 40% water and 60% air in the vadose zone, and 100% air in the atmosphere. The $CH_4$ leakage was introduced as a point source in the aquifer to simulate upward and lateral migration. Air was modeled as a pseudo-pure component, assumed insoluble under shallow subsurface conditions [48]. Fluid properties and phase behavior were calibrated using CMG's Winprop module (Supporting Information, **Section S3.1**). Section-specific relative permeability and capillary pressure functions were assigned to represent multiphase flow behavior and potential leakage to the atmosphere.

Capillary-pressure scaling for each grid block incorporated wettability variability via a contact-angle-informed (CA) Leverett J-function [49]:

$$\frac{P_{c,grid}}{P_{c,ref}} = \frac{\cos(\theta_{grid})}{\cos(\theta_{ref})} \sqrt{\frac{k_{ref}\phi_{grid}}{k_{grid}\phi_{ref}}} \qquad (1)$$

where subscripts *ref* and *grid* refer to reference conditions and model grid block values. $k$, $\phi$, $\theta$, and $P_C$ represents absolute permeability, porosity, contact angle and capillary pressure, respectively.

The first term, ($\frac{\cos(\theta_{grid})}{\cos(\theta_{ref})}$), accounts for wettability variability between facies, while the second term, ($\sqrt{\frac{k_{ref}\phi_{grid}}{k_{grid}\phi_{ref}}}$), accounts for scaling of pore-size correlation length. In the no-contact-angle (NCA) scenario, wettability is assumed constant, so the first term equals unity and the scaling reduces to the conventional porosity–permeability formulation used in most gas transport studies.

Each simulation began with a 30-day pre-leakage phase, during which injection and production wells at model boundaries operated at about 0.6 m³/day to establish steady-state groundwater flow. The $CH_4$ leakage was then introduced at a constant rate of 0.14 rm³/day for 60 days, and simulations were extended up to 100 days to ensure stabilization of leakage pathways.

The spatial distribution of CG and FG facies within both the aquifer and vadose zone was modeled using a transition-probability-based approach implemented in T-PROGS [50] (see Supporting Information, **Section S3.2**). This method generated 25 stochastic realizations of the facies architecture, each representing a distinct spatial configuration of CG and FG facies while maintaining identical volumetric proportions. The realizations differ in the geometry, spatial arrangement, and connectivity of the facies, thereby capturing geological uncertainty and allowing assessment of stochastic variability in model outcomes.

To represent different hydrogeochemical conditions, both MD and continuum-scale simulations were conducted for zero-salinity (0 ppm NaCl) and saline (12,000 ppm NaCl) groundwaters. The freshwater case represents inland aquifers recharged primarily by precipitation, while the saline case represents coastal aquifers, evaporative basins, or regions influenced by agricultural recharge [51–54].

For the base scenario, the permeability of CG facies, $CH_4$ leakage rate, and volumetric proportion of CG facies were set to 5,000 mD, 0.14 m³/day, and 40%, respectively. From this baseline, additional scenarios were developed by systematically varying key parameters: (i) increasing CG permeability from 5,000 to 20,000 mD while raising the log-scale variance from 0.3 to 1.0; (ii) increasing $CH_4$ leakage rate and total leakage volume by a factor of 2.5; (iii) increasing salinity from 0 to 12,000 ppm to assess the influence of groundwater chemistry; and (iv) increasing the volumetric proportion of CG facies from 40% to 60% to examine the effects of sedimentary variability and connectivity. The upper limit of 60% CG facies was selected because, in two-dimensional systems, the connectivity threshold occurs near 50%. The resulting $CH_4$ distributions across mobile, trapped, and dissolved phases were then compared between contact-angle included (CA) and NCA scenarios to evaluate the influence of contact angle on leakage predictions.

## 3 Results and discussion

### 3.1 Contact angle estimation

Estimated contact angles from MD simulations are presented in **Figure 1**. To confirm that simulation time was sufficient, the normalized vertical position of the water droplet center of mass was tracked for each mineral at two salinity levels (0 and 12,000 ppm NaCl). As shown in Supporting Information, **Figure S2**, all cases reached a plateau, indicating that equilibrium was achieved (6-10 and 2 ns for quartz and kaolinite, respectively). The MD-computed contact angles fall within the range of previously reported laboratory measurements for quartz (Supporting Information, **Figure S3**), supporting the reliability of the approach. Variability among experimental values likely reflects differences in sample preparation, surface conditions, measurement techniques, and operator handling. The consistency between our MD results and this experimental range further supports the validity of the modeling approach; however, the uncertainty observed in laboratory measurements should also be acknowledged and incorporated into the analysis.

Quartz is consistently more water-wet than kaolinite. Increasing salinity reduces water-wetness on both minerals, raising contact angle (θ) by about 14° on quartz (from about 20° to about 34°) and by about 3° on kaolinite (from about 63° to about 66°). The stronger salinity response on quartz is physically reasonable [56–62] as dissolved salts screen electrostatic charges at the mineral–water interface, weakening water's adhesion to the surface and enhancing gas wettability. The uncertainty in contact angle was estimated for both minerals under zero salinity and saline conditions using 10 frames from the final 1 ns of the MD simulations. For quartz, the standard deviation of the contact angle is approximately 1.7° in zero salinity groundwater and 2.6° in saline water, whereas for kaolinite, it is about 3.7° and 4.0°, respectively.

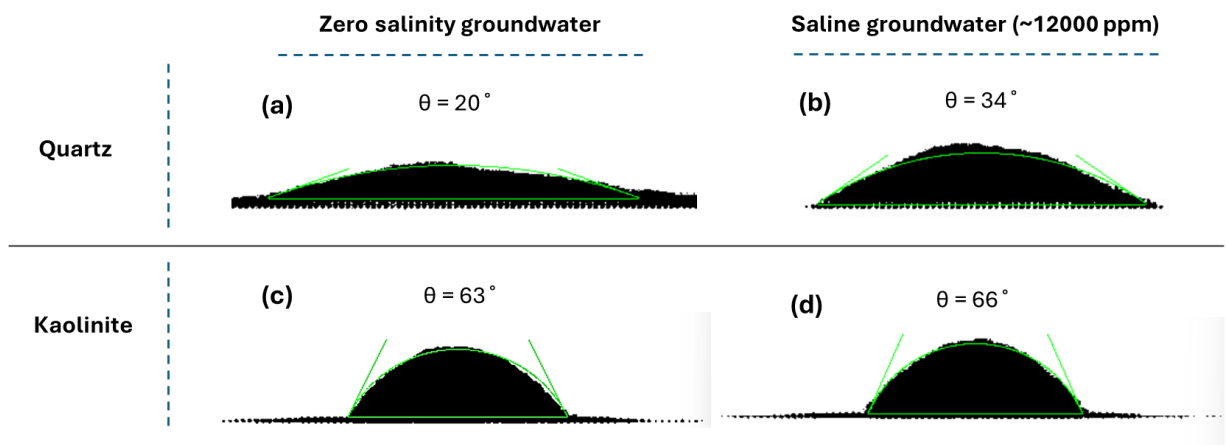

**Figure 1.** Estimated contact angle using molecular dynamics simulation based on time-averaged droplet shape for the last 2 nanoseconds during the production phase for water- $CH_4$-substrate system: (a) Zero salinity groundwater-$CH_4$-quartz, (b) Saline groundwater-$CH_4$-quartz, (c) Zero salinity groundwater-$CH_4$-kaolinite, and (d) Saline groundwater-$CH_4$-kaolinite. The light green lines indicate the boundaries of the water droplet used for contact angle determination.

## 3.2 Continuum scale simulations

After completing all simulations, we first examined how the averaged $CH_4$ (%) distribution converged as the number of realizations increased. This convergence analysis, conducted for the base scenario after 190 days of simulation, considered variations across different reservoir sections and fluid phases. The objective was to verify the statistical stability and representativeness of the simulation results, ensuring that the observed $CH_4$ migration and trapping behaviors were not influenced by random variability in mineral distributions (Supporting Information, **Section S4.1**). Once convergence was confirmed, plume propagation was compared between CA and NCA scenarios. As can be seen in Supporting Information, **Section S4.2**, including contact angle variability in scaling the capillary pressure curve (NCA) reshapes the distribution of capillary entry pressure in the reservoir by smoothing out sharp contrasts between high- and low-pressure regions, creating a new heterogeneity pattern. This redistribution reduces the lateral spread of the leaked $CH_4$ plume compared to conventional scaling (CA). Once leakage ceases, $CH_4$ movement is largely upward or becomes trapped, as capillary forces shaped by the heterogeneous entry pressures, together with gravitational forces from the density contrast between $CH_4$ and groundwater, dominate over viscous forces that would otherwise drive horizontal migration. Animations of $CH_4$ plume migration through the aquifer and vadose zone for a representative scenario were provided in **Multimedia attachments**. Results show that incorporating contact angle variability generates additional flow pathways that reduce the sharp plume fronts typically observed when only conventional capillary heterogeneity is considered. CA also leads to higher leakage to the atmosphere compared to NCA. Moreover, the animations illustrate how the mobile $CH_4$ phase in the aquifer gradually breaks into disconnected clusters once injection stops, indicating the loss of hydraulic connectivity. These isolated gas clusters become increasingly immobilized over time due to capillary trapping, which prevents further upward migration and contributes to the long-term stabilization of $CH_4$ in the subsurface.

As demonstrated, contact angles computed from MD simulations inherently contain some degree of uncertainty. Comparison with experimental measurements indicates that the MD-derived contact angles fall within the range of laboratory results; however, certain unaccounted factors may still influence wettability behavior. To account for these potential effects, a sensitivity analysis was conducted by varying the contact angle of each facies by ±10°. To capture the range of possible outcomes, continuum-scale simulations were conducted for two extreme scenarios representing maximum and minimum contrasts in facies-specific contact angles. These scenarios were used to assess the influence of contact angle uncertainty on $CH_4$ leakage predictions. Further details are provided in Supporting Information, **Section S4.3**.

**Figure 2** illustrates the uncertainty of temporal evolution of $CH_4$ partitioning across the aquifer, vadose zone, and atmosphere for 25 stochastic realizations, comparing the NCA and CA cases. These variations in facies configuration led to differences in plume migration and trapping behavior, which are reflected in the variability shown in the figure. Within the aquifer and vadose zone, $CH_4$ was categorized into mobile, snap-off trapped, and dissolved phases. As $CH_4$ quantities in the vadose zone and their associated variability were relatively small, only the total value is reported here (see Supporting Information, **Section S4.4** for details on trapped $CH_4$ in the vadose zone). In the atmosphere, only total $CH_4$ is presented, as $CH_4$ exists solely in the gaseous form in this region. In the figure, each row corresponds to a specific sensitivity scenario on base scenario such as increased permeability, elevated salinity, higher $CH_4$ leakage or increased volume proportion of CG facies, while each column shows the temporal distribution of $CH_4$ across the different sections or phases. The resulting $CH_4$ percentages represent the fraction of $CH_4$ moles in each phase or section relative to the total leaked $CH_4$ in each simulation. Variability was quantified as the standard deviation from the mean across the 25 realizations, reflecting the influence of geological heterogeneity on plume propagation and phase partitioning. In the plots, solid lines indicate mean values, shaded regions denote one standard deviation above and below the mean, and gray shaded areas highlight regions of overlap between the CA and NCA cases, illustrating where their results converge.

Within the aquifer, mobile $CH_4$ concentration increases during the leakage phase, then declines after leakage stops as $CH_4$ becomes trapped, is dissolved, or migrates upward. Among the phases, dissolved $CH_4$ shows the least variability (about 10%). Incorporating contact angle heterogeneity reduces aquifer $CH_4$ by up to 10% in both mobile and trapped phases, while atmospheric leakage increases by as much as 20%. In contrast, the vadose zone exhibits only a minor increase in $CH_4$, with negligible differences between CA and NCA averages.

At higher leakage rates, differences between CA and NCA scenarios diminish, and variability across realizations decreases. The reduction arises because viscous-dominated flow suppresses capillary effects, leading to enhanced $CH_4$ migration into the atmosphere regardless of wettability assumptions.

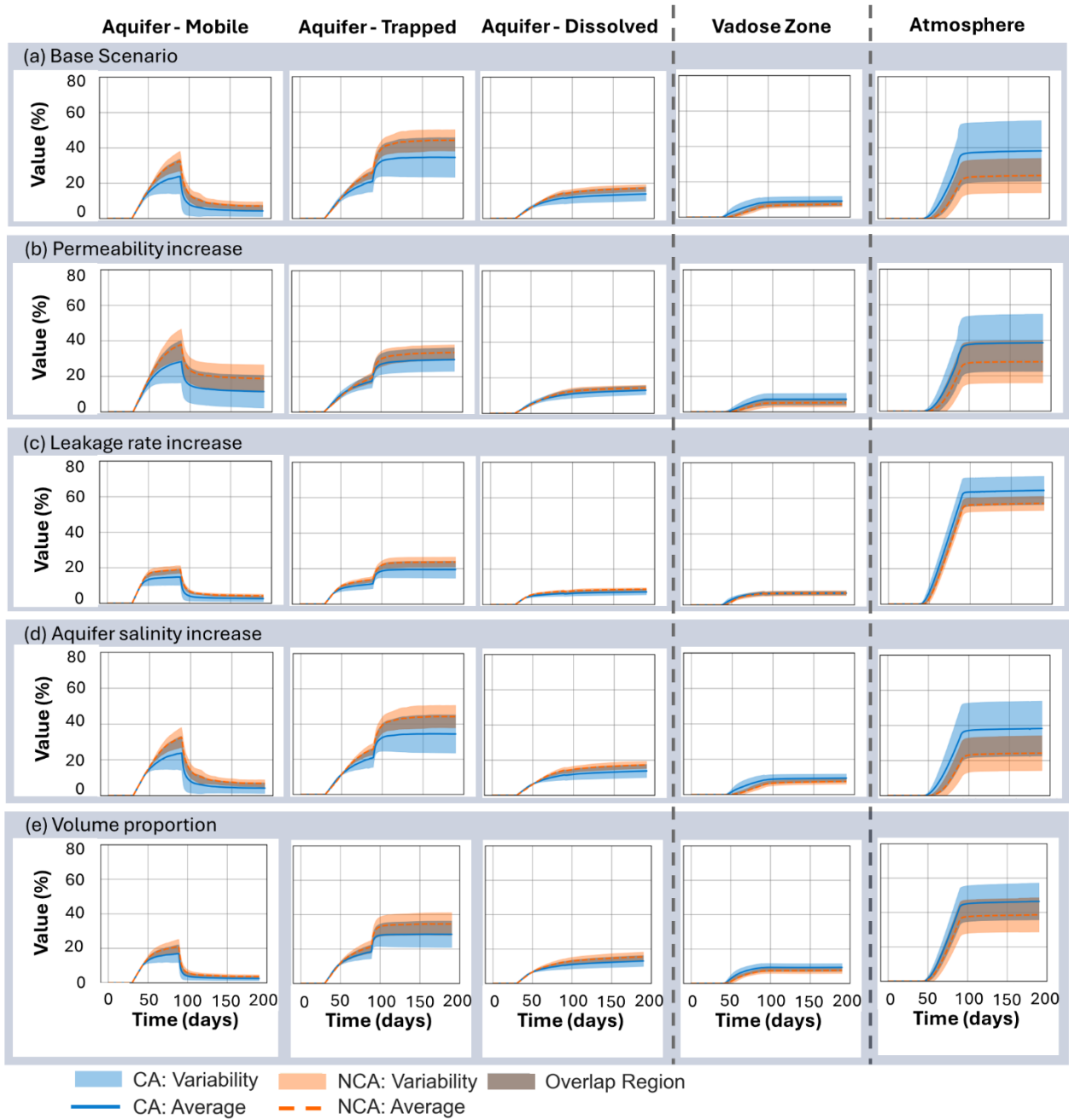

**Figure 2.** Variability of leaked $CH_4$ across 25 stochastic realizations of mineral heterogeneity. Simulations were performed in the aquifer, vadose zone, and atmosphere over 190 days, comparing cases with and without contact angle incorporation. Scenarios include: (a) Base scenario, (b) Permeability increase (4×Base), (c) Leakage rate increase (2.5×Base), (d) Aquifer salinity increase (changed from 0 to ~12,000 ppm), and (e) Volume proportion increase (coarse-grain (CG) volume proportion changed from 40% to 60%). In the aquifer, results are shown for mobile $CH_4$ gas, snap-off trapped $CH_4$, and dissolved $CH_4$; in the vadose zone and atmosphere, only total $CH_4$ gas is shown. Abbreviations: CA = contact angle included in capillary pressure scaling; NCA = no contact angle included.

While **Figure 2** illustrates variability using standard deviations symmetrically around the mean, the approach may obscure the full range of outcomes across realizations. To more explicitly capture the spread in $CH_4$ distribution, we evaluated the results after 190 days using box plots (**Figure 3**), which provide a clearer representation of variability among scenarios in each section. Each box represents the interquartile range (25th–75th percentile), with the central line indicating the median. Whiskers show the spread of non-outlier data, and individual markers denote outliers. This statistical representation highlights both the central tendency and the variability across scenarios with and without contact angle (CA versus NCA) effects.

Comparison of interquartile ranges between the CA and NCA cases indicates that incorporating contact angle enhances the variability associated with mineral heterogeneity, thereby increasing the influence of stochastic realizations. Among all scenarios, the largest variability in mobile $CH_4$ is observed under permeability increase, where the interquartile range is widest and the median difference between CA and NCA reaches nearly 10%. In contrast, scenarios involving increased leakage rate or volume proportion exhibit the smallest differences, both in variability and median values.

For snap-off trapped $CH_4$, the interquartile range generally increases when contact angle is included, except in the permeability increase case. Median values exhibit the largest CA–NCA difference (about 20%) in the base case and in the salinity increase scenario, although the absolute values differ. This suggests that aquifer salinity, which can vary with environmental conditions, influences $CH_4$ partitioning across phases. Notably, despite the variations, the difference between CA and NCA remains consistent in the trapped phase, indicating that contact angle effects must be incorporated for accurately quantifying $CH_4$ trapping. Dissolved $CH_4$ follows a trend similar to trapped $CH_4$.

In the vadose zone, CA scenarios consistently yield slightly higher mobile $CH_4$ fractions compared to NCA, with median differences below 3%. The exception is the leakage rate increase scenario, where both median and variability are nearly identical. $CH_4$ percentages in this section range between 9–12% for CA, reflecting the absence of capillary trapping due to loose sediments and the transitional role of the vadose zone between aquifer and atmosphere.

Atmospheric leakage is the most critical factor for environmental impact. As expected, scenarios with increased leakage rate and volume result in the greatest $CH_4$ release to the atmosphere, with median values of approximately 75% for CA and 70% for NCA. These findings highlight the importance of incorporating contact angle variability into modeling efforts, as it alters $CH_4$ partitioning among phases and directly affects the magnitude of emissions reaching the atmosphere.

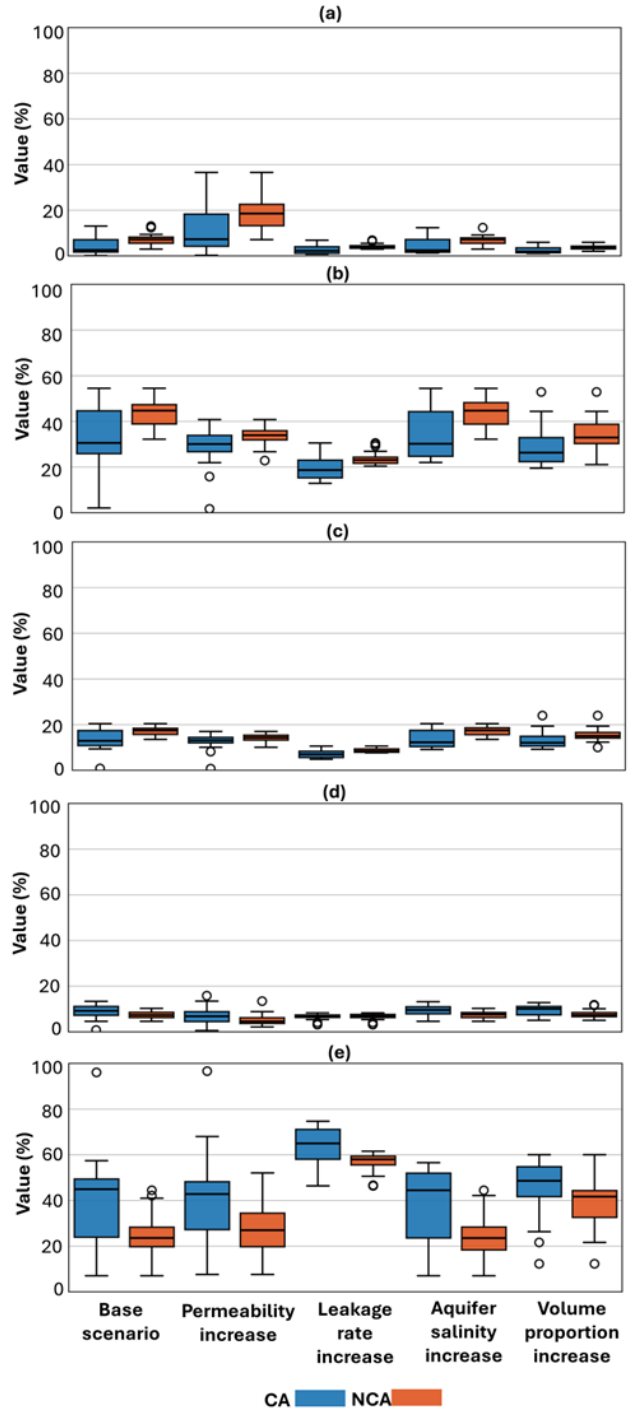

**Figure 3.** Percentage of leaked $CH_4$ across the aquifer, vadose zone, and atmosphere after 190 days, when the $CH_4$ plume has stabilized in different scenarios: Base scenario, permeability increase, leakage rate increase and volume proportion increase. Results are shown for scenarios with and without contact angle incorporation. (a) mobile $CH_4$ gas in aquifer, (b) snap-off trapped $CH_4$ in aquifer, (c) dissolved $CH_4$ in aquifer; (d) total $CH_4$ gas in the vadose zone; (e) total $CH_4$ gas

in the atmosphere. CA = contact angle included in capillary pressure scaling; NCA = contact angle not included; CG = coarse-grain rock type.

## 4   Implications and future research

Our results extend prior field-calibrated evidence that capillary heterogeneity exerts a significant effect on $CH_4$ plume geometry and fate in shallow aquifers. Whereas simplifying or ignoring capillary heterogeneity leads to overestimated atmospheric $CH_4$ and underestimated mobile $CH_4$ in the aquifer [15]. We demonstrate that heterogeneous wettability, expressed through spatially variable water–$CH_4$ contact angles as a function of mineralogy and salinity, adds an independent and quantifiable effect by directly modifying local capillary entry pressures even when permeability and porosity are fixed. Incorporating contact-angle variability (CA) changes partitioning: aquifer mobile and snap-off trapped $CH_4$ decrease by up to 10%, while atmospheric leakage increase by up to 20%. The CA–NCA difference decreases at high leakage rates, consistent with a transition toward viscous-dominated flow.

Our findings have important practical implications for leakage risk assessment. Conventional models that assume constant wettability (NCA) underestimate atmospheric leakage in settings where mineralogy and groundwater salinity covary, such as quartz-rich coastal or evaporite aquifers [16]. This bias arises because salinity increases the contact angle on quartz much more than on kaolinite, lowering capillary entry pressures in quartz-dominated saline systems and promoting earlier gas breakthrough and higher leakage flux. Conversely, decreasing groundwater salinity can reduce the contact angle and raise capillary entry pressures, thus, enhancing gas trapping and limiting leakage. To quantify these dynamics, a contact-angle-informed Leverett $J$-function, Equation (1), should replace conventional scaling that accounts only for porosity and permeability, particularly in mixed-facies aquifers with sharp mineralogical contrasts or evolving salinity. This approach complements the findings of Ershadnia et al. (2021) [16], which highlighted the dominant role of meter-scale facies architecture in controlling plume migration. They demonstrated that heterogeneous wettability acts at the same scales and must therefore be jointly characterized, mapped, and modeled to improve predictions of subsurface gas transport and leakage risk.

Recent advances in micro-CT imaging have enabled direct extraction of contact-angle fields from pore-scale images, demonstrating that wettability is inherently spatially heterogeneous. rather than taking on a single uniform value [63]. Building on this evidence, a practical approach for applying CT results at the field scale is to use CT data as a facies-level calibrator. In this workflow, a limited number of representative core plugs from each facies can be imaged, and their contact-angle distributions computed. The resulting statistics can then be mapped as a facies-conditioned contact-angle field throughout the reservoir model. Our contact-angle–informed *J*-function is designed to incorporate such fields directly, enabling propagation of pore-scale heterogeneity into continuum-scale flow simulations. Where CT data are unavailable, MD simulation-computed contact-angles serve as physically grounded alternative that can later be refined as even limited CT measurements become available. This hybrid strategy circumvents the challenges of direct CT-to-field upscaling, ties wettability estimates to measured rock textures, and naturally facilitates uncertainty quantification through ensemble sampling alongside variations in permeability contrast, leakage rates, and salinity.

In this study, we focused on clean quartz and kaolinite surfaces and assumed static contact angles; however, natural sediments often exhibit additional complexities such as surface roughness (asperities), organic coatings, iron oxide precipitates, and contact angle hysteresis (advancing vs. receding angles), all of which can significantly influence wettability and plume migration. Future experimental and modeling efforts should address the effect of these factors to better represent realistic leakage in unconfined shallow aquifers. Moreover, salinity was held constant in our simulations, whereas in nature, groundwater chemistry evolves over time due to changing boundary conditions, which in turn can alter contact angle and capillary behavior. Continuum-scale measurements of contact angle variability and leaked $CH_4$ plume propagation across space and time are therefore crucial to validate and refine predictive models and to capture the dynamics of $CH_4$ leakage. Such data, combined with improved models, will enhance the accuracy of risk assessments for groundwater quality and fugitive $CH_4$ emissions.

The implications of our findings extend beyond shallow aquifers. Heterogeneous wettability remains a critical unknown in deep geologic systems, including $CO_2$ and $H_2$ storage or toxic chemicals such as NAPL, where high mobility ratios make capillary heterogeneity, a key factor controlling plume propagation and contributing to short-term trapping [19,20,23,27,28,64]. Addressing such knowledge gaps by integrating laboratory, field, with multiscale modeling studies will improve predictive capabilities for both shallow and deep subsurface fluid migration and support the development of safer and more effective energy and climate mitigation technologies.

## 5 Abbreviations

| Abbreviation | Explanation |
|---|---|
| CA, NCA | Contact angle included, excluded scenarios |
| FG, CG | Fine, coarse grain |
| MD | Molecular dynamics |

| Chemical formula | Explanation |
|---|---|
| $CH_4$ | Methane |
| $CO_2$ | Carbon dioxide |
| $H_2$ | Hydrogen |

| Parameter | Explanation |
|---|---|
| $k$ | Permeability in mD |
| $\phi$ | Porosity in fraction |
| $P_c$ | Capillary pressure |
| $j(S_w)$ | J-function as a function of water saturation |
| $\sigma$ | Interfacial tension |
| $\theta$ | Contact angle |

| Units | Explanation |
|---|---|
| Bar | Metric unit of pressure |
| nm | Nanometer |
| mD | Milli-Darcy |
| g/mol | Gram per mole |
| K | Kelvin |
| ppm | Part per million |
| rm³/day | Cubic meter per day at aquifer condition |

## 6 Data availability

The simulation samples from both molecular dynamics and continuum-scale models generated in this study are publicly available on Zenodo [https://doi.org/10.5281/zenodo.17459998], enabling access and reproducibility of the results.

## 7 Acknowledgment

The authors acknowledge financial support from the American Chemical Society Petroleum Research Fund under the New Directions Grant (ACS PRF #67019-ND2). We gratefully thank the Ohio Supercomputer Center for providing computational resources and expertise essential to this work. We also thank the Computer Modelling Group for generously providing CMG software to the Department of Geosciences at the University of Cincinnati.